\documentclass[runningheads]{llncs}

\usepackage[T1]{fontenc}
\usepackage[utf8]{inputenc}
\usepackage{textcomp} 
\usepackage{graphicx}
\usepackage{tikz}
\usetikzlibrary{fit}
\usepackage{array}
\usepackage{xurl}
\usepackage{placeins}

\DeclareUnicodeCharacter{00A0}{~}
\DeclareUnicodeCharacter{2011}{-}
\DeclareUnicodeCharacter{2013}{--}
\DeclareUnicodeCharacter{2014}{---}
\DeclareUnicodeCharacter{2018}{`}
\DeclareUnicodeCharacter{2019}{'}
\DeclareUnicodeCharacter{201C}{``}
\DeclareUnicodeCharacter{201D}{''}

\newcommand{\citep}[1]{\cite{#1}}
\newcommand{\citet}[1]{\cite{#1}}
\begin{document}
\usetikzlibrary{fit}

\title{Outer Limits: An Experimental Approach to Controlled Content Manipulation within the Reddit Interface}
\titlerunning{Outer Limits for Controlled Content Experiments}

\author{
Chenchen Mao\inst{1}\and
Hanjing Shi\inst{1}\and
Haiyan Jia\inst{1}\and
Daniel Unguryan\inst{1}\and
Eric Baumer\inst{2}\and
Dominic DiFranzo\inst{1}
}

\authorrunning{Mao et al.}

\institute{
Lehigh University, Bethlehem, PA, USA
\email{\{chm321, hasa23, haj616, djd219\}@lehigh.edu}
\and
University of Toronto, Toronto, ON, Canada\\
\email{eps.baumer@utoronto.ca}
\and
Computer Science \& Engineering, Lehigh University, Bethlehem, PA, USA\\
\email{unguryan77@gmail.com}
}

\maketitle

\begin{abstract}
Independent researchers often lack access to intervention capabilities for
controlled experiments on live social media platforms. We present Outer Limits, a browser-based system for controlled content
experiments within the existing Old Reddit interface, rather than in a
reconstructed simulation. The system renders content locally,
records study events, and contains configured voting and commenting actions so
that neither constructed content nor experimental write interactions reach
Reddit. In a 219-participant perceptual-fidelity study, ART ANOVAs found no
significant Post Type, Participant Awareness, or interaction effects.
Exploratory TOSTs met the $d=\pm0.50$ equivalence criterion for the marginal
contrasts and for Post Type within the forewarned subgroup. We also illustrate
the system with a factorial study varying post frame, comment frame, and
comment stance. Outer Limits combines three properties that the approaches
considered here provide separately: precise control over experimental
content, an existing platform interface, and containment of experimental
content and interactions from the host community.
\end{abstract}

\section{Introduction}

Understanding user behavior on social media informs research on political
polarization, misinformation, and public health
\cite{Conover_Ratkiewicz_Francisco_Goncalves_Menczer_Flammini_2021,doi:10.1126/science.aao2998,10.1145/3544548.3580946,charles2015using}.
These problems require identifying which platform features and content change
behavior, not simply documenting what users do. Controlled experiments can help
identify such causal relationships \cite{gilbert_handbook_1998}, including how
perceived group norms shape responses to online misinformation
\cite{10.1145/3544548.3580946}.

When a hypothesized mechanism depends on a platform's social and interface
context, conducting the experiment within that platform allows the effect to
be estimated in the setting in which the mechanism operates. Platform-run
studies, such as Meta's controversial ``emotional contagion'' experiment,
illustrate this form of experimentation
\cite{doi:10.1073/pnas.1320040111}. Such interventions, however,
have required internal access or platform cooperation, which independent
researchers rarely have.

Independent researchers therefore rely on observational data, which lacks
randomized control
\cite{bruns_after_2019,Mozilla_2021,Bechmann_Vahlstrup_2015,10.1145/2827872},
or on vignettes and simulations, which provide control but do not reproduce the
social embeddedness of the target platform
\cite{doi:10.1027/1614-2241/a000014,vraga_addressing_2021,JAGAYAT2024101726}:
participants' actions are not visible to their real networks and carry no real
reputational consequences. Findings from a reconstructed environment therefore
leave open whether the same effect would emerge on the platform itself.

Experiments conducted within the platform can reduce this uncertainty. Recent
client-side field experiments show that browser extensions can do
this without platform cooperation, by reranking the content participants
encounter in their own feeds \cite{piccardi_reranking_2026,piccardi_reranking_2025}. Because that content is drawn from each participant's personalized feed, however,
the specific posts and comments differ across participants. Independent
researchers therefore still lack a way to present precisely matched,
researcher-specified content within an existing interface while keeping that
content and participants' interactions from reaching the host community.

We present an experimental approach to controlled content manipulation within
an existing social media interface, instantiated in Outer Limits. The approach
links platform selection, localized content manipulation, containment, and
fidelity validation. Outer Limits implements matched content variation, event
logging, and containment of configured write interactions. We evaluate
perceptual fidelity with 219 participants and illustrate its use in a factorial
experiment.

The contribution is limited to content-level interventions in the studied
Reddit setting. The validation applies only to the tested outcomes,
equivalence bounds, materials, and sample.

\section{Related Work}

\subsection{Controlled Experiments in Reconstructed Social Media Environments}

Independent researchers frequently use reconstructed environments to conduct
controlled social media experiments without relying on platform cooperation.
Static vignette studies present participants with hypothetical scenarios, such
as text descriptions or screenshots of social media posts
\cite{doi:10.1027/1614-2241/a000014,vraga_addressing_2021}.
These approaches allow researchers to manipulate content while holding other
stimulus features constant, but static stimuli offer limited interaction with
the surrounding platform environment. This may constrain how participants engage
with the material compared with a live platform.

More interactive alternatives include the (Mis)information Game
\cite{butler_mis_2024}, the Mock Social Media Website Tool
\cite{jagayat2021mock}, and the Truman Platform
\cite{10.1145/3173574.3173785}. These systems combine experimental control
with an interactive interface that participants can browse, react to, and
engage with.

The main limitation is not necessarily perceived artificiality, because
participants may find these environments engaging. A reconstructed environment,
however, does not reproduce the social embeddedness of the target platform.
Participants' actions are not
visible to their actual networks and carry none of the reputational
consequences of a real like, comment, or share. When anticipated audiences and
platform conventions shape how participants interpret content and decide whether
to engage, findings from a reconstructed environment leave open whether the same
effect would emerge on the target platform.

\subsection{Independent Experiments on Existing Platforms}

An alternative is to intervene within an existing platform interface rather
than reconstructing it.  Historically this has depended on
platform-controlled infrastructure: Meta's
``emotional contagion'' study altered the news feeds of nearly 700,000 users
\cite{doi:10.1073/pnas.1320040111}, while a 61-million-person Facebook
experiment examined how messages containing social cues affected voter turnout
\cite{bond_61-million-person_2012}. These studies demonstrate the scale and
directness of platform-based experimentation but generally required platform
access or cooperation.

Independent researchers have often relied instead on observational data, even
as platforms have increasingly restricted API access
\cite{bruns_after_2019,davidson2023platform,ying_2025}. User-donation
approaches, including Mozilla's RegretsReporter, Digital Footprints, and
MovieLens, partially address these access barriers
\cite{Mozilla_2021,Bechmann_Vahlstrup_2015,10.1145/2827872}. These approaches
provide access to naturally occurring behavior but do not by themselves
provide randomized manipulation of content or interface features. Causal
identification therefore requires additional assumptions or research designs.

Client-side reranking experiments provide a more direct route to independent
intervention within existing platforms. Piccardi et al.
describe a browser-extension method that intercepts and reranks existing
content in participants' social media feeds without platform cooperation and
apply this method in a field experiment
\cite{piccardi_reranking_2025,piccardi_reranking_2026}. Because that content comes from each participant's own feed, however, the specific posts and comments differ across
participants, and their wording remains outside the researcher's control.

Intervening within a live platform also carries a cost that reconstructed
environments do not: experimental actions can affect users or communities that
did not consent to participate. The r/ChangeMyView moderation team reported an
unauthorized experiment that introduced undisclosed AI-generated comments without
prior community consultation \cite{cmv_mod_team_2025}. These concerns motivate a
central design requirement of Outer Limits: constructed content is not published
to Reddit, and designated participant interactions do not generate corresponding
platform write requests. This containment addresses spillover into the host
platform and community, though it does not by itself resolve questions of
deception, consent, or debriefing.

\subsection{Positioning Outer Limits}

These approaches differ in the interface participants use, what the researcher
controls, the social context participants perceive, and what reaches the platform
(Table~\ref{tab:design-space}). Vignettes and simulations give the researcher
full control over content, but the environment is built for the study and
participants' actions reach no real audience. Client-side reranking keeps
participants on the existing platform and controls which posts appear, but not
their wording. Outer Limits controls what the posts say, presents them within the
existing Reddit community, and prevents constructed content and participant
interactions from reaching the platform.
\begin{table}[t]
\centering
\small
\setlength{\tabcolsep}{4pt}
\renewcommand{\arraystretch}{1.2}
\begin{tabular}{
>{\raggedright\arraybackslash}p{1.9cm}
>{\raggedright\arraybackslash}p{2.05cm}
>{\raggedright\arraybackslash}p{2.2cm}
>{\raggedright\arraybackslash}p{2.0cm}
>{\raggedright\arraybackslash}p{1.95cm}}
\hline
\textbf{Approach} &
\textbf{Interface} &
\textbf{What is controlled} &
\textbf{Social context} &
\textbf{Platform effects} \\
\hline
Static vignette &
Static image or text &
Displayed content &
None &
None \\
Standalone simulation &
Reconstructed &
Content and features &
Known to be constructed &
None \\
Client-side reranking &
Existing feed &
Exposure and ordering &
Participant's own network &
Real \\
Outer Limits &
Existing Reddit &
Matched content variants &
Existing community &
Contained \\
\hline
\end{tabular}
\caption{Design space for social media experiments. The table compares design
characteristics, not validated outcomes, and does not imply a hierarchy of
methods.}
\label{tab:design-space}
\end{table}

Outer Limits is designed for studies requiring this combination. Standalone
simulations or client-side reranking experiments may remain preferable when a
study prioritizes, respectively, complete control over the full experimental
environment or naturalistic exposure to existing personalized feeds and ordinary
downstream platform behavior.
\section{Design Principles}
\label{sec:existing-platform}

Outer Limits begins with an existing platform interface rather than a
reconstructed simulation. This design is useful when interpretation of the
treatment depends on
platform context, while the causal comparison requires matched,
researcher-specified content. Starting from an existing interface does not
imply naturalistic recruitment, unrestricted platform use, or a field
experiment.

The approach rests on four principles.

\begin{description}

\item[\textbf{Platform Selection:}]
Researchers select a platform whose interface matters to the research question
and can be modified in the participant's browser. This approach is most useful
when posts, comments, or other presentation cues are part of the experimental
context.

\item[\textbf{Localized Content Manipulation:}]
Researchers change only the posts, comments, or other elements required by the
experimental design, while keeping the surrounding interface and
nonmanipulated content unchanged. This allows precise and consistent
manipulation without access to the platform's servers.

    \item[\textbf{Containment:}]
Constructed content and participant actions such as votes and comments should
remain within the study and should not reach or affect the host platform or
its community.

    \item[\textbf{Fidelity Validation:}]
Manipulated content should appear naturally integrated into the host
interface. Building on an existing platform preserves its established
mechanics but limits how much researchers can customize
\cite{10.1145/3555557}. Inconsistencies may reveal the intervention and change
how participants interpret the content. Covert manipulations therefore require
evidence that altered content does not create substantial differences on the
selected fidelity measures; disclosed interventions need not meet the same
standard. Perceptual fidelity contributes to experimental realism, but does
not establish it in full. Participants need only find the situation credible
enough to take it seriously and engage with the content as intended
\cite{aghajari_methodological_2023}.
\end{description}

\section{Outer Limits: Design and Implementation}
\label{outer}

Outer Limits implements this approach through a Chrome extension and backend for
controlled content experiments within the existing Reddit interface. We selected
Reddit because its structure suits social media experiments. Its division into
thousands of topic-specific communities (``subreddits'') offers substantial
diversity for research questions \cite{10.1145/2492517.2492646}, and the platform
accommodates content types from simple text to rich media. Reddit also has an
established culture of ``throwaway accounts'' for discussing sensitive topics
\cite{10.1145/2675133.2675175}, making accounts without long-standing identities
familiar on the platform.

\subsection{System Architecture and Data Flow}

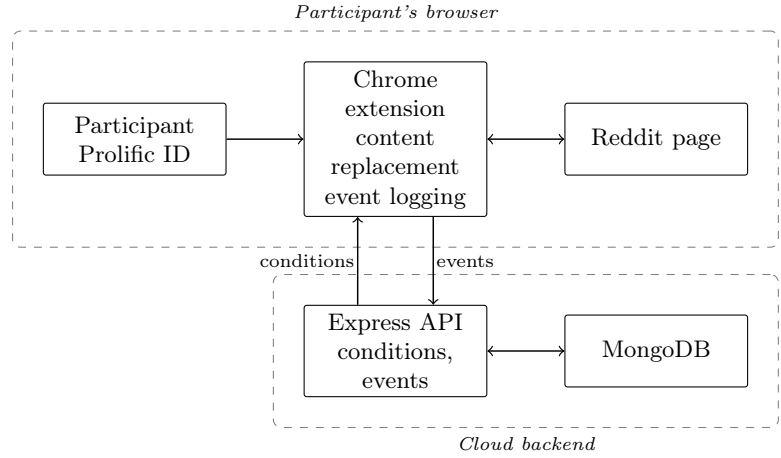
\begin{figure}[!h]
\centering
\begin{tikzpicture}[
    x=1cm, y=1cm,
    box/.style={draw, rounded corners=1pt, align=center, minimum height=0.95cm, text width=2.2cm, font=\footnotesize},
    flow/.style={->, line width=0.55pt},
    note/.style={font=\scriptsize, fill=white, inner sep=1pt},
    group/.style={draw, dashed, rounded corners=3pt, inner sep=4mm, gray}
]
\node[box] (participant) at (0,1.45) {Participant\\Prolific ID};
\node[box] (extension) at (3.45,1.45) {Chrome extension\\content replacement\\event logging};
\node[box] (reddit) at (6.9,1.45) {Reddit page};
\node[box] (api) at (3.45,-1.35) {Express API\\conditions, events};
\node[box] (storage) at (6.9,-1.35) {MongoDB};

% dashed group boxes
\node[group, fit=(participant)(extension)(reddit), label={[font=\scriptsize\itshape]above:Participant's browser}] {};
\node[group, fit=(api)(storage), label={[font=\scriptsize\itshape]below:Cloud backend}] {};

\draw[flow] (participant) -- (extension);
\draw[<->, line width=0.55pt] (extension) -- (reddit);
\draw[flow] ([xshift=-5mm]api.north) -- node[note, left] {conditions} ([xshift=-5mm]extension.south);
\draw[flow] ([xshift=5mm]extension.south) -- node[note, right] {events} ([xshift=5mm]api.north);
\draw[<->, line width=0.55pt] (api) -- (storage);
\end{tikzpicture}
\caption{Outer Limits architecture. The extension fetches the assigned condition
from the backend, replaces content in the Reddit page, and logs interactions.
Constructed content and participants' votes and comments stay within the
participant's browser and the study backend; they are not sent to Reddit.}
\label{fig:outer-workflow}
\end{figure}

Outer Limits consists of two components (Figure~\ref{fig:outer-workflow}).

\textbf{The Chrome extension} runs in the participant's browser. It fetches the
participant's assigned experimental condition from the backend, replaces the
designated content in the Reddit page, logs the participant's interactions, and
prevents those interactions from reaching Reddit.

\textbf{The backend} is an Express API with a MongoDB database. It stores the
experimental conditions that researchers configure, assigns participants to
conditions, and receives the interaction records the extension returns.

A session begins when the participant enters a randomly generated Prolific
identifier, which serves as a pseudonym in the study backend. The backend either
retrieves an existing assignment for that identifier or
assigns the participant to a condition, balancing enrollment across conditions.
The identifier subsequently links the participant's condition assignment to their
interaction records. Uninstalling the extension ends its ability to modify pages
or record interactions.

\FloatBarrier

\subsection{Researcher Configuration and Deployment}

Researchers specify experiments through two CSV files. A post file defines each
target URL and the content to be displayed there: title, body text, image,
displayed author, timestamp, vote count, and the condition to which the post
belongs. A comment file links comments to posts and specifies their text,
displayed author, timestamp, and vote count. An upload script writes these records
to the study database.

Because the extension reads all experimental content from the database at runtime,
researchers can revise stimuli, add conditions, or change a factorial design
without modifying or rebuilding the extension. We hosted the backend on Render and
used MongoDB Atlas for storage; both are implementation choices rather than
requirements of the approach.

For participant deployment, we published the extension through the Chrome Web
Store so that participants could install it directly rather than manually load
an unpacked extension. This distribution route was intended to reduce participant
setup steps and the installation support required during study enrollment. The
listing address is omitted for anonymous review.
\subsection{Runtime Manipulation and Containment}
\paragraph{Content replacement.}
When a participant opens a target page, the extension requests the assigned
condition from the backend and uses the returned fields to replace the
corresponding elements in the page DOM: the post's title, body text, image,
displayed author, timestamp, and vote count, together with the comment thread. The
surrounding interface (the subreddit header, sidebar, navigation, and any
nonmanipulated content) continues to load from Reddit, so the intervention remains
localized to the experimental content unit. Constructed content is rendered only in
the participant's browser and is never published to Reddit.

\paragraph{Interaction logging.}
The extension logs designated study interactions: post views, upvotes, downvotes,
comment submissions, and subsequent comment edits. Each record carries the
participant identifier, a timestamp, the action type, and the relevant post or
comment metadata, and is sent to the backend.

\paragraph{Containment of write interactions.}
Logging alone does not prevent an interaction from reaching Reddit; containment is
implemented separately for each write action. For votes on manipulated content, the
extension intercepts the participant's click before Reddit's own event handler
executes, prevents the event from propagating to that handler, updates the
displayed vote state locally, and submits an interaction record to the backend. For
comment submissions, the extension captures the submitted text and stores it as a
study record without generating the corresponding write request to Reddit. From the
participant's perspective, the interface responds to voting and commenting, but
no corresponding action reaches Reddit's servers.

\paragraph{Software availability.}
Outer Limits is open-source. Its repository includes the Chrome extension, the
backend, an upload script, example CSV files, and a step-by-step deployment
guide. The address is omitted for anonymous review.

\section{Validation: Assessing Perceptual Fidelity}

The validation tested whether participants perceived manipulated posts as realistic
as authentic posts and whether forewarning them about possible alterations changed
those perceptions. Forewarning served as a boundary condition: if ratings remained
similar when participants expected alterations, casual browsing would be less
likely to explain the result.

\noindent\textbf{RQ1.} Do participants perceive manipulated posts as realistic as
authentic posts within the existing Reddit interface?

\noindent\textbf{RQ2.} Does explicitly forewarning participants that content may
have been altered change these perceptions?

\subsubsection{Method}
We conducted a validation study on Prolific using a $2 \times 2$
between-subjects factorial design. The factors were \textbf{Post Type}
(authentic vs. manipulated) and \textbf{Participant Awareness} (uninformed vs.
informed).

\begin{figure}[!t]
\centering
\begin{minipage}[t]{0.49\textwidth}
\centering
\textbf{(a) Authentic image-based post}\par\smallskip
\includegraphics[width=\linewidth,trim=0 250 250 0,clip]{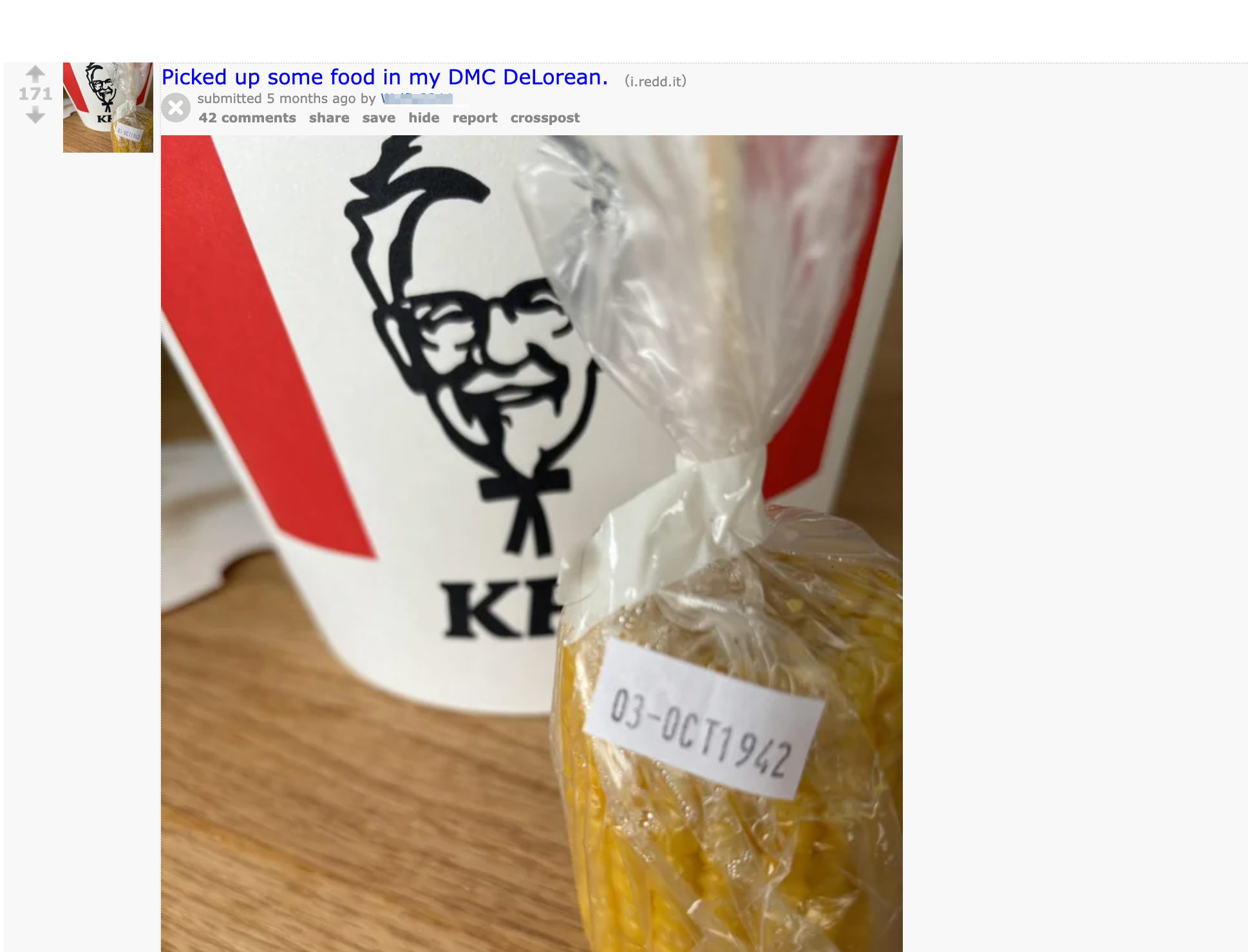}
\end{minipage}\hfill
\begin{minipage}[t]{0.49\textwidth}
\centering
\textbf{(b) Manipulated image-based post}\par\smallskip
\includegraphics[width=\linewidth,trim=0 330 100 0,clip]{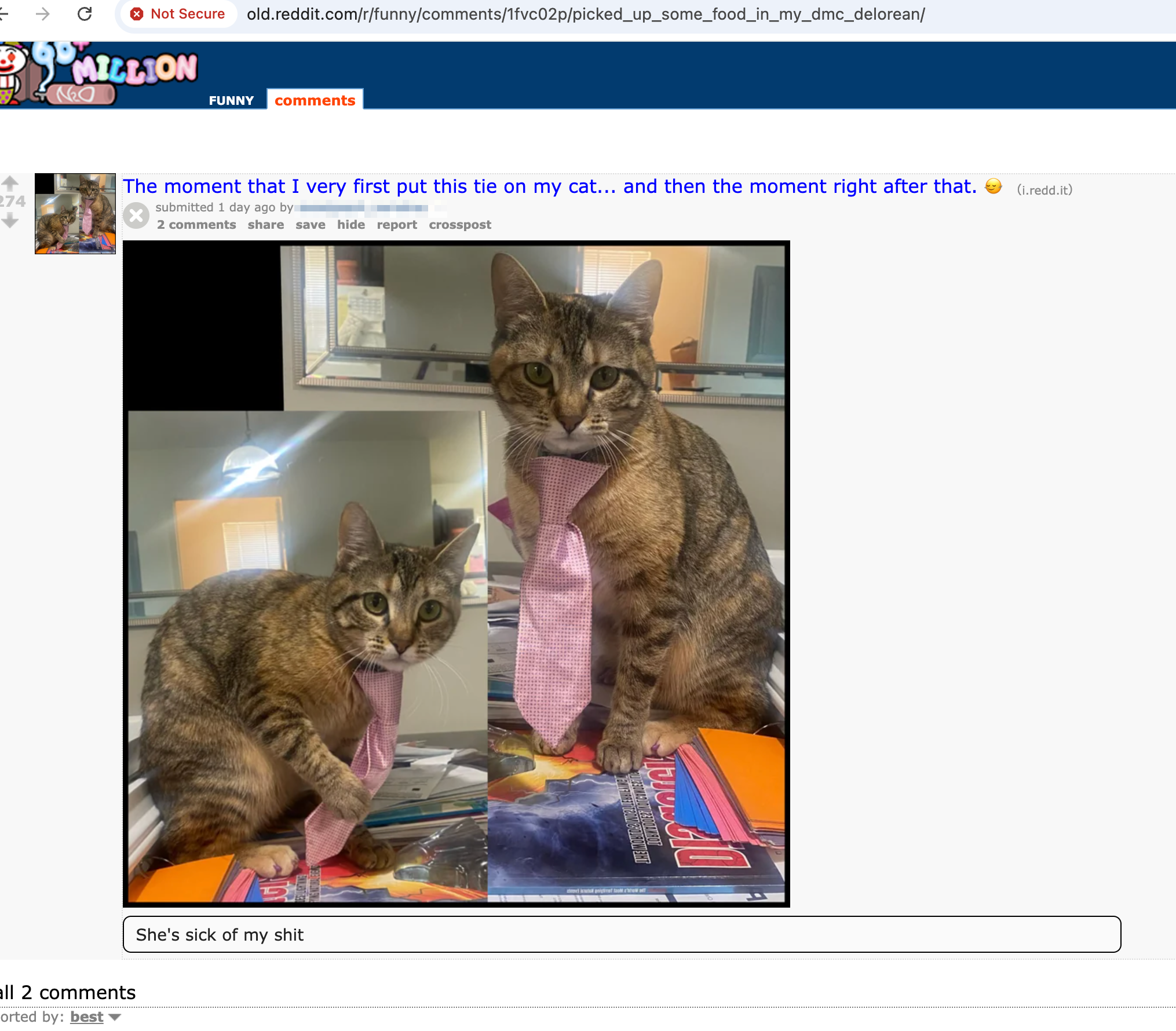}
\end{minipage}
\caption{Image-based validation stimuli shown at a readable scale. Outer Limits retained the target URL and surrounding Reddit interface while replacing the title, post text, image, and comments in panel (b).}
\label{fig:validation-image}
\end{figure}

\begin{figure}[!t]
\centering
\textbf{(a) Authentic text-only post}\\[-1mm]
\includegraphics[width=0.92\textwidth]{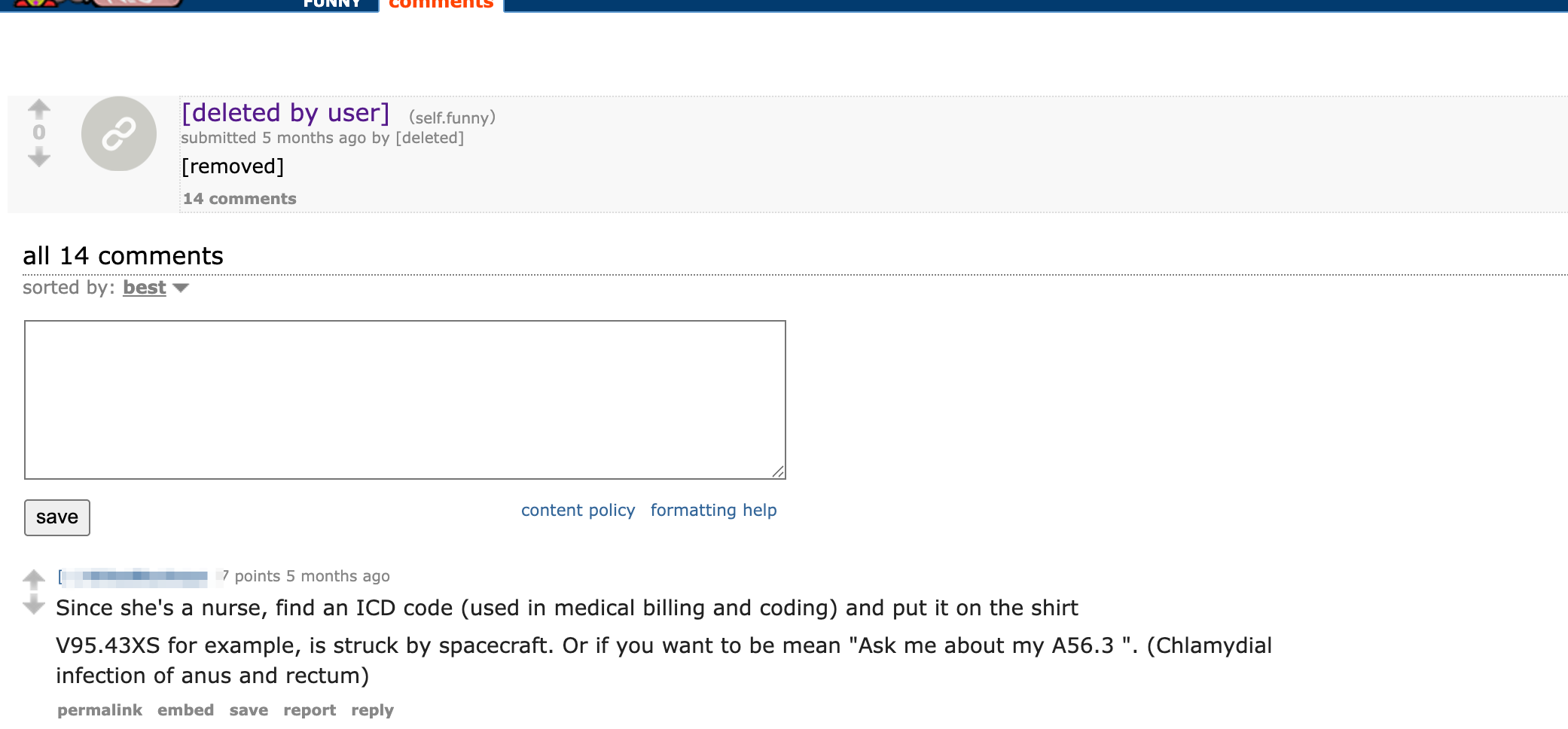}

\vspace{2mm}

\textbf{(b) Manipulated text-only post}\\[-1mm]
\includegraphics[width=0.92\textwidth,trim=0 150 0 0,clip]{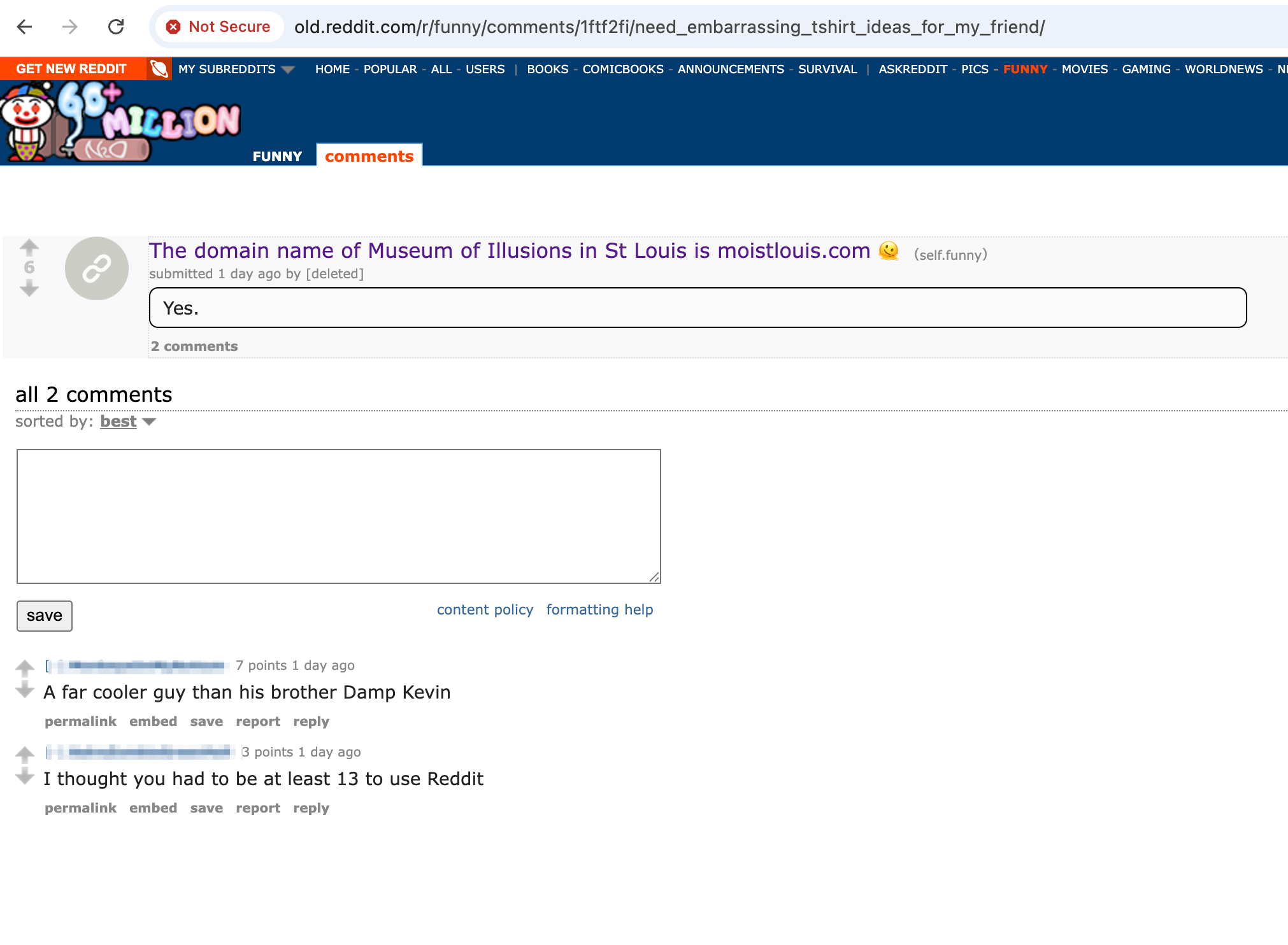}
\caption{Text-only validation stimuli shown at a readable scale. The authentic post's title and body had been deleted by its owner. Outer Limits retained the target URL and surrounding Reddit interface while replacing the title, post text, and comments in panel (b).}
\label{fig:validation-text}
\end{figure}

For study materials, we selected one image-based and one text-only post from
Reddit's \texttt{r/funny} community. In the experimental condition, Outer Limits
replaced their title, text, image, and comments with content from the same
subreddit. The URL remained unchanged (Figures~\ref{fig:validation-image}
and~\ref{fig:validation-text}).

After a pre-survey, participants installed the extension and entered their
Prolific identifier. The backend assigned them to one of four conditions.
Participants in the informed condition were told that the posts might have been
altered. All participants viewed the two target links and completed a post-survey
containing content-recall checks and perceptual-fidelity measures, followed by a
debriefing. The study received IRB approval at [Anonymous] University (protocol
\#2171134-1).

\subsubsection{Variables and Measurements} \label{questions}

Participants rated four retrospective statements on a 7-point scale
($1=$ Strongly Disagree, $7=$ Strongly Agree): the way interactive elements
(e.g., upvote/downvote buttons and comment placement) functioned seemed
different; the overall look and layout of the posts or page appeared different
from usual; the content of the posts, including text, images, and comments,
seemed altered; and ``I did not notice anything unusual.'' They also rated
confidence from ``not confident at all'' to ``very confident'' and described
anything unusual in an open-ended response.

\subsubsection{Participants}
Between November and December 2024, we recruited 225 participants via Prolific.
Six participants were removed for failing condition-specific content-recall
checks, which required correctly identifying the content shown in their assigned
condition; 219 participants remained. Compensation was \$6.50. An a priori power analysis ($f = 0.25$, power
= .95) required $N = 210$. Mean age was 33.9 years (SD = 9.9); 63\% identified as male and 33\% as female.
Demographics did not differ by condition (all $p > .18$).

\FloatBarrier
\subsection{Validation Results}

\subsubsection{Quantitative Analysis}

\begin{table}[t]
\centering
\footnotesize
\setlength{\tabcolsep}{2.2pt}
\renewcommand{\arraystretch}{1.08}

\begin{tabular}{@{}lccccc@{}}
\hline
& \multicolumn{3}{c}{\textbf{ART ANOVA: $F$ ($p$)}}
& \multicolumn{2}{c}{\textbf{TOST ($d=\pm0.50$)}} \\
\cline{2-4}\cline{5-6}

\textbf{Response}
& \shortstack{\textbf{Post}\\\textbf{Type}}
& \textbf{Awareness}
& \textbf{Interaction}
& \shortstack{\textbf{Marginal}\\\textbf{Post/Aware.}}
& \shortstack{\textbf{Post within}\\\textbf{forewarned}} \\
\hline

Interaction
& 1.53 (.217)
& 0.05 (.830)
& 0.39 (.531)
& Eq./Eq.
& Eq. \\

Layout
& 1.34 (.249)
& 0.78 (.378)
& 0.65 (.422)
& Eq./Eq.
& Eq. \\

Content
& 1.98 (.161)
& 0.01 (.904)
& 0.37 (.545)
& Eq./Eq.
& Eq. \\

\shortstack[l]{Nothing\\unusual}
& 2.07 (.152)
& 0.04 (.841)
& 0.08 (.774)
& Eq./Eq.
& Eq. \\

Confidence
& 0.08 (.771)
& 0.49 (.483)
& 0.07 (.790)
& Eq./Eq.
& Eq. \\

\hline
\end{tabular}

\caption{ART ANOVA and exploratory TOST results. ART cells report $F$ ($p$);
all effects were nonsignificant. Eq.\ denotes that both one-sided tests rejected
effects at or beyond $d=\pm0.50$ at $\alpha=.05$ for the marginal Post
Type/Awareness contrasts and Post Type within the forewarned subgroup ($n=58$
vs.\ 49).}

\label{tab:realism-tests}
\end{table}

We used ART ANOVAs to test Post Type, Participant Awareness, and their
interaction. Because nonsignificant difference tests do not establish
equivalence, we also conducted exploratory TOSTs using $d=\pm0.50$ bounds
\cite{lakens2017}. The TOSTs examined the two marginal contrasts and Post Type
within the forewarned subgroup; the bounds were not specified in advance.

No ART effect was significant (Table~\ref{tab:realism-tests}). All reported TOST contrasts met the
$d=\pm0.50$ equivalence criterion, supporting statistical equivalence within
these bounds for the tested outcomes and setting. This does not establish zero
difference or experimental realism in full.

\subsubsection{Qualitative Analysis}
Two authors calibrated the coding scheme on an initial 10\% of responses and
independently coded a second 10\%, reaching Krippendorff's $\alpha > 0.8$ across
seven themes: No Changes Noticed, Layout Issues, Content Suspicion, Clear
Alterations, Uncertain, Technical Issues, and Other. They then independently
coded the full dataset and resolved remaining disagreements through discussion.
We fitted a Poisson regression to theme-by-condition counts with theme category,
Post Type, Participant Awareness, and their interaction as predictors. Neither
Post Type $(\beta = -0.03, p = .910)$, Participant Awareness $(\beta = -0.11,
p = .647)$, nor their interaction $(\beta = 0.35, p = .259)$ predicted theme
frequencies. The qualitative results were consistent with the quantitative
results.

\FloatBarrier
\section{Illustrative Case Study: A Factorial Reddit Study}
\label{sec:case_study}

We illustrate Outer Limits with a factorial study of how post frame, comment
frame, and comment stance shape health beliefs and behavioral intentions
\cite{10.1145/3544548.3580946,janz_health_1984}. Direct deployment would
require publishing vaccine-related experimental content, while a standalone
simulation would remove the Reddit context.

The $2\times2\times2$ design varied post frame, comment frame, and comment
stance while holding usernames, timestamps, metrics, and other page elements
constant. Participants completed a pre-survey, installed the extension, viewed
their condition, and completed a post-survey followed by debriefing. The study
received IRB approval (protocol \#2171134-2). We report the design rather than
substantive outcomes.

\FloatBarrier

\section{Discussion}
\label{sec:discussion}

\subsection{Contribution}

Outer Limits combines matched content manipulation within the existing Reddit
interface with containment of constructed content and configured write
interactions. By separating the interface participants experience from the
platform consequences their actions would otherwise produce, it supports studies
in which platform context matters but unrestricted downstream behavior is
unnecessary. For computational social science, this configuration makes matched
content comparisons possible when platform APIs do not provide intervention
access. It also allows researchers to study sensitive content without publishing
it to host communities.

Methodologically, this approach complements observational platform studies and
reconstructed experiments rather than replacing either. It allows
theory-derived content mechanisms to be tested under randomized, matched
conditions while retaining the interface cues through which those mechanisms
may operate. This combination can help researchers examine whether effects
attributed to content depend on the interface context in which participants
encounter it. Evidence from such experiments could also inform assumptions used
in computational models, although the present study does not test that use.

\subsection{Relation to Existing Approaches}

Compared with reconstructed environments, Outer Limits retains the host interface
instead of requiring researchers to reproduce selected platform conventions, but
it offers less control over the surrounding environment and remains vulnerable to
platform changes \cite{JAGAYAT2024101726}. Piccardi et al.\ likewise use a browser extension to intervene without platform
cooperation \cite{piccardi_reranking_2025,piccardi_reranking_2026}, but they
rerank the content participants encounter in their own feeds rather than
specifying it. Outer Limits instead assigns matched, researcher-specified posts
and comments, records designated events, and contains configured write
interactions.
This combination is useful when exact content contrasts, existing platform cues,
and bounded platform consequences all matter, but not when naturalistic feed use
is required.

\subsection{Ethical Considerations}

Containment is both a technical and an ethical boundary. It prevents constructed
content and configured interactions from reaching the host community, but it does
not eliminate the ethical questions that covert manipulation raises. Participants
act under the belief that their votes and comments carry ordinary platform
consequences; in fact these interactions are contained. The same belief may
contribute to behavioral credibility but also constitutes deception. We
addressed it through IRB review and full debriefing, and researchers deploying
this approach should do the same, in addition to documenting and testing which
write interactions are intercepted and protecting backend data.

\subsection{Limitations}

The current implementation targets desktop Old Reddit and does not support newer
Reddit interfaces, mobile clients, or video content. Because content replacement
depends on the rendered page structure, changes to Reddit's DOM can require
updates and renewed fidelity assessment. The validation itself covers two posts
from one subreddit, evaluated by a Prolific sample; the equivalence criterion was
met using bounds selected after data collection. Fidelity is a property of particular
materials and settings, not an automatic consequence of technical integration, so
each new deployment requires its own assessment.

\subsection{Future Work}

The most direct test of whether the existing platform interface affects an
estimated treatment effect would compare the same manipulation in Outer Limits
and a reconstructed simulation. The approach could also support matched studies
of message framing, moderation cues, social
norms, and misinformation responses in settings where community context is part of
the treatment. Adaptive computational social science experiments could vary content
in real time according to prespecified participant actions. Such extensions would
require auditable adaptation rules, controls against
unintended treatment variation, and renewed review of consent, deception, and
containment. Applications to other interfaces or platforms would require renewed
technical, fidelity, and ethical review in each setting.
\section{Conclusion}
This paper presents an experimental approach to controlled content
manipulation within an existing social media interface and implements it in
Outer Limits for Old Reddit. A 219-participant study
supports perceptual fidelity within the tested equivalence bounds and setting.
An illustrative factorial case applies Outer Limits to matched post and comment
variations. Relative to the reconstructed simulations and feed-reranking methods
examined here, Outer Limits combines researcher-specified content, the host
interface, and bounded downstream consequences in one experimental design.

\bibliographystyle{splncs04}
\bibliography{references}

@article{lakens2017,
	title = {Equivalence Tests: A Practical Primer for t Tests, Correlations, and Meta-Analyses},
	author = {Lakens, Dani\"el},
	journal = {Social Psychological and Personality Science},
	year = {2017},
	volume = {8},
	number = {4},
	pages = {355--362},
	doi = {10.1177/1948550617697177},
}

@misc{cmv_mod_team_2025,
	title = {Unauthorized Experiment on {CMV} Involving {AI}-Generated Comments},
	author = {{r/ChangeMyView Moderation Team}},
	month = apr,
	year = {2025},
	url = {https://www.reddit.com/r/changemyview/comments/1k8b2hj/meta_unauthorized_experiment_on_cmv_involving/},
	note = {Reddit community announcement, accessed 13 July 2026},
}

@article{piccardi_reranking_2025,
	title = {Reranking partisan animosity in algorithmic social media feeds alters affective polarization},
	volume = {390},
	doi = {10.1126/science.adu5584},
	number = {6776},
	urldate = {2026-07-12},
	journal = {Science},
	publisher = {American Association for the Advancement of Science},
	author = {Piccardi, Tiziano and Saveski, Martin and Jia, Chenyan and Hancock, Jeffrey and Tsai, Jeanne L. and Bernstein, Michael S.},
	month = nov,
	year = {2025},
	pages = {eadu5584},
}

@article{piccardi_reranking_2026,
	title = {Reranking {Social} {Media} {Feeds}: {A} {Practical} {Guide} for {Field} {Experiments}},
	volume = {9},
	shorttitle = {Reranking {Social} {Media} {Feeds}},
	doi = {10.1145/3800557},
	number = {1},
	urldate = {2026-07-12},
	journal = {ACM Transactions on Social Computing},
	author = {Piccardi, Tiziano and Saveski, Martin and Jia, Chenyan and Hancock, Jeffrey T. and Tsai, Jeanne and Bernstein, Michael},
	month = mar,
	year = {2026},
	pages = {2:1--2:17},
}

@article{janz_health_1984,
	title = {The {Health} {Belief} {Model}: a decade later},
	volume = {11},
	issn = {0195-8402},
	shorttitle = {The {Health} {Belief} {Model}},
	doi = {10.1177/109019818401100101},
	language = {eng},
	number = {1},
	journal = {Health Education Quarterly},
	author = {Janz, N. K. and Becker, M. H.},
	year = {1984},
	pages = {1--47},
}

@article{aghajari_methodological_2023,
	title = {Methodological {Middle} {Spaces}: {Addressing} the {Need} for {Methodological} {Innovation} to {Achieve} {Simultaneous} {Realism}, {Control}, and {Scalability} in {Experimental} {Studies} of {AI}-{Mediated} {Communication}},
	volume = {7},
	shorttitle = {Methodological {Middle} {Spaces}},
	doi = {10.1145/3579506},
	number = {CSCW1},
	urldate = {2025-10-05},
	journal = {Proc. ACM Hum.-Comput. Interact.},
	author = {Aghajari, Zhila and Baumer, Eric P. S. and Hohenstein, Jess and Jung, Malte F. and DiFranzo, Dominic},
	month = apr,
	year = {2023},
	pages = {73:1--73:28},
}

@article{bond_61-million-person_2012,
	title = {A 61-million-person experiment in social influence and political mobilization},
	volume = {489},
	copyright = {2012 Springer Nature Limited},
	issn = {1476-4687},
	doi = {10.1038/nature11421},
	language = {en},
	number = {7415},
	urldate = {2025-10-04},
	journal = {Nature},
	publisher = {Nature Publishing Group},
	author = {Bond, Robert M. and Fariss, Christopher J. and Jones, Jason J. and Kramer, Adam D. I. and Marlow, Cameron and Settle, Jaime E. and Fowler, James H.},
	month = sep,
	year = {2012},
	pages = {295--298},
}

@article{vraga_addressing_2021,
	title = {Addressing {COVID}-19 {Misinformation} on {Social} {Media} {Preemptively} and {Responsively}},
	volume = {27},
	issn = {1080-6040},
	doi = {10.3201/eid2702.203139},
	number = {2},
	urldate = {2025-10-04},
	journal = {Emerging Infectious Diseases},
	author = {Vraga, Emily K. and Bode, Leticia},
	month = feb,
	year = {2021},
	pages = {396--403},
}

@book{gilbert_handbook_1998,
	title = {The {Handbook} of {Social} {Psychology}},
	isbn = {978-0-19-521376-8},
	language = {en},
	publisher = {McGraw-Hill},
	author = {Gilbert, Daniel Todd and Fiske, Susan T. and Lindzey, Gardner},
	year = {1998},
}

@article{butler_mis_2024,
	title = {The (mis) information game: {A} social media simulator},
	volume = {56},
	number = {3},
	journal = {Behavior Research Methods},
	publisher = {Springer},
	author = {Butler, Lucy H and Lamont, Padraig and Wan, Dean Law Yim and Prike, Toby and Nasim, Mehwish and Walker, Bradley and Fay, Nicolas and Ecker, Ullrich KH},
	year = {2024},
	pages = {2376--2397},
}

@article{bruns_after_2019,
	title = {After the ‘{APIcalypse}’: social media platforms and their fight against critical scholarly research},
	volume = {22},
	doi = {10.1080/1369118X.2019.1637447},
	number = {11},
	journal = {Information, Communication \& Society},
	publisher = {Routledge},
	author = {Bruns, Axel},
	year = {2019},
	pages = {1544--1566},
}

@article{JAGAYAT2024101726,
	title = {A primer on open-source, experimental social media simulation software: {Opportunities} for misinformation research and beyond},
	volume = {55},
	issn = {2352-250X},
	doi = {10.1016/j.copsyc.2023.101726},
	journal = {Current Opinion in Psychology},
	author = {Jagayat, Arvin and Choma, Becky L.},
	year = {2024},
	pages = {101726},
}

@inproceedings{10.1145/3173574.3173785,
	address = {Montreal QC, Canada},
	series = {Chi '18},
	title = {Upstanding by design: {Bystander} intervention in cyberbullying},
	isbn = {978-1-4503-5620-6},
	doi = {10.1145/3173574.3173785},
	booktitle = {Proceedings of the 2018 {CHI} conference on human factors in computing systems},
	publisher = {Association for Computing Machinery},
	author = {DiFranzo, Dominic and Taylor, Samuel Hardman and Kazerooni, Franccesca and Wherry, Olivia D. and Bazarova, Natalya N.},
	year = {2018},
	pages = {1--12},
}

@inproceedings{10.1145/2492517.2492646,
	address = {Niagara, Ontario, Canada},
	series = {Asonam '13},
	title = {An exploration of discussion threads in social news sites: a case study of the {Reddit} community},
	isbn = {978-1-4503-2240-9},
	doi = {10.1145/2492517.2492646},
	booktitle = {Proceedings of the 2013 {IEEE}/{ACM} international conference on advances in social networks analysis and mining},
	publisher = {Association for Computing Machinery},
	author = {Weninger, Tim and Zhu, Xihao Avi and Han, Jiawei},
	year = {2013},
	pages = {579--583},
}

@inproceedings{10.1145/2675133.2675175,
	address = {Vancouver, BC, Canada},
	series = {Cscw '15},
	title = {"{This} is a throwaway account": {Temporary} technical identities and perceptions of anonymity in a massive online community},
	isbn = {978-1-4503-2922-4},
	doi = {10.1145/2675133.2675175},
	booktitle = {Proceedings of the 18th {ACM} conference on computer supported cooperative work \& social computing},
	publisher = {Association for Computing Machinery},
	author = {Leavitt, Alex},
	year = {2015},
	pages = {317--327},
}

@article{doi:10.1126/science.aao2998,
	title = {The science of fake news},
	volume = {359},
	doi = {10.1126/science.aao2998},
	number = {6380},
	journal = {Science},
	author = {Lazer, David M. J. and Baum, Matthew A. and Benkler, Yochai and Berinsky, Adam J. and Greenhill, Kelly M. and Menczer, Filippo and Metzger, Miriam J. and Nyhan, Brendan and Pennycook, Gordon and Rothschild, David and Schudson, Michael and Sloman, Steven A. and Sunstein, Cass R. and Thorson, Emily A. and Watts, Duncan J. and Zittrain, Jonathan L.},
	year = {2018},
	pages = {1094--1096},
}

@article{doi:10.1073/pnas.1320040111,
	title = {Experimental evidence of massive-scale emotional contagion through social networks},
	volume = {111},
	doi = {10.1073/pnas.1320040111},
	number = {24},
	journal = {Proceedings of the National Academy of Sciences},
	author = {Kramer, Adam D. I. and Guillory, Jamie E. and Hancock, Jeffrey T.},
	year = {2014},
	pages = {8788--8790},
}

@article{doi:10.1027/1614-2241/a000014,
	title = {Experimental vignette studies in survey research},
	volume = {6},
	doi = {10.1027/1614-2241/a000014},
	number = {3},
	journal = {Methodology},
	author = {Atzmüller, Christiane and Steiner, Peter M.},
	year = {2010},
	pages = {128--138},
}

@misc{Mozilla_2021,
	title = {Mozilla investigation: {YouTube} algorithm recommends videos that violate the platform’s very own policies},
	url = {https://foundation.mozilla.org/en/blog/mozilla-investigation-youtube-algorithm-recommends-videos-that-violate-the-platforms-very-own-policies/},
	author = {{Mozilla}},
	month = jul,
	year = {2021},
}

@article{10.1145/2827872,
	title = {The {MovieLens} datasets: {History} and context},
	volume = {5},
	issn = {2160-6455},
	doi = {10.1145/2827872},
	number = {4},
	journal = {ACM Transactions on Interactive Intelligent Systems},
	publisher = {Association for Computing Machinery},
	author = {Harper, F. Maxwell and Konstan, Joseph A.},
	month = dec,
	year = {2015},
}

@inproceedings{10.1145/3544548.3580946,
	address = {Hamburg, Germany},
	series = {Chi '23},
	title = {What’s the norm around here? {Individuals}’ responses can mitigate the effects of misinformation prevalence in shaping perceptions of a community},
	isbn = {978-1-4503-9421-5},
	doi = {10.1145/3544548.3580946},
	booktitle = {Proceedings of the 2023 {CHI} conference on human factors in computing systems},
	publisher = {Association for Computing Machinery},
	author = {Aghajari, Zhila and Baumer, Eric P. S. and DiFranzo, Dominic},
	year = {2023},
}

@article{Bechmann_Vahlstrup_2015,
	title = {Studying facebook and instagram data: {The} digital footprints software},
	volume = {20},
	doi = {10.5210/fm.v20i12.5968},
	number = {12},
	journal = {First Monday},
	author = {Bechmann, Anja and Vahlstrup, Peter Bjerregaard},
	month = dec,
	year = {2015},
}

@article{Conover_Ratkiewicz_Francisco_Goncalves_Menczer_Flammini_2021,
	title = {Political polarization on twitter},
	volume = {5},
	doi = {10.1609/icwsm.v5i1.14126},
	number = {1},
	journal = {Proceedings of the International AAAI Conference on Web and Social Media},
	author = {Conover, Michael and Ratkiewicz, Jacob and Francisco, Matthew and Goncalves, Bruno and Menczer, Filippo and Flammini, Alessandro},
	month = aug,
	year = {2021},
	pages = {89--96},
}

@article{charles2015using,
	title = {Using social media for actionable disease surveillance and outbreak management: a systematic literature review},
	volume = {10},
	number = {10},
	journal = {PloS one},
	publisher = {Public Library of Science San Francisco, CA USA},
	author = {Charles-Smith, Lauren E and Reynolds, Tera L and Cameron, Mark A and Conway, Mike and Lau, Eric HY and Olsen, Jennifer M and Pavlin, Julie A and Shigematsu, Mika and Streichert, Laura C and Suda, Katie J and {others}},
	year = {2015},
	pages = {e0139701},
}

@misc{ying_2025,
	title = {The evolving landscape of web scraping on social media platforms {\textbar} d-lab},
	url = {https://dlab.berkeley.edu/news/evolving-landscape-web-scraping-social-media-platforms},
	author = {Ying, Nanqin},
	month = mar,
	year = {2025},
}

@article{davidson2023platform,
	title = {Platform-controlled social media {APIs} threaten open science},
	volume = {7},
	number = {12},
	journal = {Nature Human Behaviour},
	publisher = {Nature Publishing Group UK London},
	author = {Davidson, Brittany I and Wischerath, Darja and Racek, Daniel and Parry, Douglas A and Godwin, Emily and Hinds, Joanne and Van Der Linden, Dirk and Roscoe, Jonathan F and Ayravainen, Laura and Cork, Alicia G},
	year = {2023},
	pages = {2054--2057},
}

@article{10.1145/3555557,
	title = {Revisiting piggyback prototyping: {Examining} benefits and tradeoffs in extending existing social computing systems},
	volume = {6},
	doi = {10.1145/3555557},
	number = {CSCW2},
	journal = {Proc. ACM Hum.-Comput. Interact.},
	publisher = {Association for Computing Machinery},
	author = {Epstein, Daniel A. and Liu, Fannie and Monroy-Hernández, Andrés and Wang, Dennis},
	month = nov,
	year = {2022},
}

@article{jagayat2021mock,
	title = {Mock social media website tool (1.0)},
	journal = {Computer software.[accessed 2024 Sep 10]. https://docs. studysocial. media},
	author = {Jagayat, Arvin and Boparai, Gurkaran and Pun, Carson and Choma, BL},
	year = {2021},
}

\end{document}